\documentclass[11pt,twoside]{article}
\usepackage{cozumel2005}
\usepackage{epsf}
\usepackage{psfig}
\usepackage{lscape}
\def\dist{(m-M)_0}
\def\logz{\lbrack\hbox{M/H}\rbrack}

\newcommand{\mean}[1]{\langle #1 \rangle}

\begin{document}
\title{Star Formation Histories of Local Group Dwarf Galaxies}    
\author{Andrew E. Dolphin}   
\affil{Steward Observatory, University of Arizona, 933 North Cherry Avenue, Tucson, AZ 85721; adolphin@as.arizona.edu}    
\author{Daniel R. Weisz, Evan D. Skillman}
\affil{Department of Astronomy, University of Minnesota, 116 Church Street SE, Minneapolis, MN 55455}
\author{Jon A. Holtzman}
\affil{Astronomy Department, New Mexico State University, Box 30001, MSC 4500, Las Cruces, NM 88003}

\begin{abstract} 
We report preliminary results of a project to homogeneously measure star formation histories for the majority of Local Group dwarf galaxies, based on archival WFPC2 observations.  For the dwarf spheroidal and elliptical galaxies, we find that the more luminous systems all show extended star formation episodes, while the less luminous systems typically formed most of their stars over ten Gyr ago.  Irregular galaxies show no comparable trend.  We find that a galaxy's classification as an irregular or spheroidal is entirely explained by its star formation within the past Gyr or less.  Several elliptical/spheroidal systems actively formed stars within the past two Gyr, and would have been indistinguishable from present-day irregulars if observed two Gyr ago.
\end{abstract}


\section{Introduction}

The Local Group is home to a diverse collection of dwarf galaxies, which show wide ranges in mass, luminosity, and stellar content.  While the dwarf galaxies comprise only a small fraction of the Local Group's total mass and light, they are astrophysically important for a variety of reasons, such as their role as laboratories in which star formation processes can be studied in the absence of spiral density waves and interactions.

Dwarf galaxies fall roughly into two morphological classes: spheroidal and irregular.  As the names suggest, spheroidal systems have fairly round and smooth contours, while irregular systems tend to be dominated by bright regions of current or very recent star formation.  Several other trends are related to the star formation/morphological type relationship: irregular galaxies have $H \alpha$ emission and high gas fractions, while spheroidals do not.  Additionally, one finds that the older populations of irregular galaxies tend to show smoother, rounder contours similar to those of spheroidals \citep[for example,][]{mill01}.

What is unclear is the cause of the irregular/spheroidal differences.  One clue may be found in the spatial distribution of dwarfs within the Local Group, shown in Figure \ref{fig_LG}.  Spheroidals tend to be found near the massive galaxies (M31 and the Galaxy), while irregulars tend to be isolated.  Another clue may exist in the presence of a small number of systems whose properties are between those of spheroidal and irregular galaxies and thus do not fall neatly into the classification scheme.  However, all of these clues are based on the present-day conditions of the systems.  To understand why systems show their current properties, we need to know their characteristics in the past -- something we can achieve by measuring their star formation histories.

\begin{figure}
\caption{The Local Group neighborhood, centered on the Galaxy.  Large black circles represent the spirals, red circles the ellipticals and spheroidals, blue triangles the irregulars, and orange diamonds the transition-type galaxies.  The grid spacing is 250 kpc.  To avoid clutter, names of systems less than 300 kpc from M31 or the Galaxy have been omitted.  Note that color is present only in the electronic version. \label{fig_LG}}
\end{figure}

During the past decade, advances in stellar evolution models and computing power have facilitated increasingly precise methods for measuring star formation histories \citep[for example,][]{tol96,dol97,apar97}, but these have generally been carried out in a sufficiently inhomogeneous fashion that comparisons are difficult.  To fill this need, we have begun work on a legacy HST program designed to measure star formation histories of the majority of Local Group systems.  Our photometry was taken from the HST Local Group stellar photometry archive (Holtzman, Afonso, \& Dolphin in prep), which contains WFPC2 photometry of all Local Group galaxies with broadband colors.  We present preliminary results of our history program here.

With the star formation histories in hand, one can begin addressing galaxy morphologies by examining the variety of star formation histories within a morphological class, as well as the similarities and differences between the morphological classes.

\section{Measuring Star Formation Histories}

\subsection{Procedure}

Before examining the star formation histories of Local Group dwarfs, it is useful to review how these histories are measured.  Regardless of the details of the measurement procedure, the underlying approach is constant: to find the star formation history most likely to have produced the observed data.  In principle, this approach can be used with any type of data on a galaxy -- variable star content, integrated colors or spectra, color-magnitude diagrams (CMDs), stellar line indices, etc.  In practice, the best data for the largest number of stars are contained in CMDs, so the history measurement process becomes one of modeling the observed CMD using synthetic CMDs corresponding to various histories.

In order to carry out this measurement, one needs the ability to create a synthetic CMD for any potential star formation history, as well as a way to determine how well each synthetic CMD matches the observed data.  The procedure for doing this was described in detail by \citet{dol02} and will not be repeated here.

We have used the isochrones of \citet{gir00} to generate our star formation histories, as these are the only set that presently have sufficiently high-mass stars to model the youngest populations, the necessary metallicity range, and complete evolution (including the horizontal branch and TP-AGB phases).  Other sets of models \citep{van00,piet04} lack high-mass stars but could be used for galaxies without ongoing star formation; however to produce a homogeneous set of histories we prefer to use a single set of models throughout.

\subsection{Accuracy of Measurement}

To test the accuracy of the measurement process, we have created a star formation history based loosely on what we measured for the Tucana dwarf.  Using this history and the artificial star library from the Tucana photometry, we have created fifty random realizations of the CMD and measured a star formation history, distance, and foreground extinction of each.

\begin{figure}
\caption{Results of our Monte Carlo tests.  The left panel shows star formation rate vs. age; the right panel shows the age-metallicity relations.  In both panels, the input values are shown by the red dotted lines.  The solid blue lines with error bars show the recovered mean values and rms scatter. \label{fig_MC}}
\end{figure}

Figure \ref{fig_MC} shows the mean recovered star formation history and age-metallicity relation, along with the rms scatter.  As is apparent from these plots, the input relations were indeed recovered correctly.  In addition, the distance modulus and $V$ extinction were recovered correctly with rms scatter of $0.006$ magnitudes.

An additional point is worth mentioning.  To produce the most useful star formation history, one would generally prefer a binning scheme for which the uncertainties in SFR(t) are relatively constant, and it is apparent from Figure \ref{fig_MC} that the logarithmic scheme adopted here is close in this regard.

\subsection{Isochrone Uncertainties}

\begin{figure}
\caption{Residuals between the observed CMD and the best-fitting model.  At left is the COZ7 data set from the Cozumel experiment (Holtzman et al., this volume); as expected the residual pattern is almost completely random.  The residuals against an actual CMD, shown at right for IC 1613, are mostly random but show coherent structure in the red clump region. \label{fig_res}}
\end{figure}

While the tests in the preceding section address the statistical accuracy of the solution, they fail to demonstrate how accurately a history can be measured, given that there are uncertainties in the isochrones.  For the majority of cases, the CMD fit will be fairly good, in that the difference between the observed and synthetic CMD is mostly noise.  However, it is common to see coherent patterns in the red clump and supergiant regions in even the best fits (Figure \ref{fig_res}), indicating minor isochrone errors.  Making fits with small offsets in distance and extinction will simulate such errors, and the rms scatter between those fits is a reasonable proxy for the resulting uncertainties.

The most significant problem exists in the case of a young system (most stars $< 2-3$ Gyr), which will contain a large number of bright red clump stars.  Unfortunately, this is an evolutionary phase that is particularly poorly modeled, and a distance measurement using CMD fitting will produce a distance that is a few tenths of a magnitude too short \citep{dol03}.  This can be used as a diagnostic.  If one finds those numbers being drawn away from reasonable values, it is useful to produce three solutions: the original fit (at the correct distance and extinction), a second using a coarser CMD binning, and a third in which the offending evolutionary phase is cut out.  Variations between the three solutions give an idea of how much the fit is affected by the isochrone errors; it is typically surprisingly little.

Thus, despite the imperfections of isochrones, a carefully-designed star formation history routine will produce accurate results provided that care is taken.

\section{Ellipticals and Spheroidals}

Armed with a general understanding of how star formation histories are measured, we now turn to the dwarf galaxies of the Local Group.  The first set of galaxies to be examined are the dwarf ellipticals and dwarf spheroidals, which are generally assumed to have the simplest star formation histories because of the lack of current or very recent star formation.

Unless otherwise noted, the CMDs are taken from the HST Local Group stellar photometry archive (Holtzman et al. in prep) and the star formation histories have been measured by us.

\subsection{Sculptor}

\begin{figure}[h]
\caption{CMD \citep[][from CTIO 4m observations]{hkmg99} and star formation history \citep{hk00} for the Sculptor dSph. For this and subsequent plots, star formation rates are given in units of $M_\odot$ Myr$^{-1}$. Star formation histories are shown graphically using population boxes \citep{hod89}. \label{fig_Sculptor}}
\end{figure}

We begin this discussion with an examination of five representative early-type systems.  The first such galaxy is the Sculptor dwarf, whose CMD and star formation history are shown in Figure \ref{fig_Sculptor}.

The CMD of this system is typical of the low-mass dwarf spheroidals.  One finds a relatively narrow main sequence turnoff corresponding to ancient stars and a well-populated horizontal branch.  Indeed, the CMD is not unlike that of an old globular cluster.  There are a non-zero number of stars above the ancient main sequence turnoff; whether these are blue stragglers or a small number of younger stars is unknown.

Not surprisingly, the star formation history measured by \citet{hk00} is that of an ancient system.  Over 90\% of the stars formed older than 10 Gyr ago, and the small amount of more recent star formation is due to the stars that may or may not be blue stragglers.  (If they are blue stragglers, the entire population of Sculptor could be ancient.)

\subsection{M110 (NGC 205)}

\begin{figure}[h]
\caption{CMD and star formation history for M110. In this and subsequent plots, isochrones are taken from \citet{gir00} and correspond to ages of $10^7$ (blue in the electronic version), $10^8$ (green), $10^9$ (orange), and $10^10$ (red) years. \label{fig_NGC205}}
\end{figure}

The second representative system is M110, which falls at the luminous end of the Local Group early-type galaxies.  Its CMD and star formation history are shown in Figure \ref{fig_NGC205}.  Comparing its CMD with that of Sculptor, there are several striking differences.  First is the presence of a dominant red clump and only a weak horizontal branch, which falls near the turnoff of the 1 Gyr isochrone.  (That the slight overdensity of stars in this part of the CMD is a horizontal branch rather than a young turnoff or noise is demonstrated by the presence of RR Lyraes \citep{saha92}.)  The strength of the red clump relative to the horizontal branch indicates that the bulk of star formation happened much more recently than in Sculptor.  Indeed, the stars bright and blue of the red clump are likely blue loop stars with ages from several hundred Myr to $\sim 1$ Gyr.

In addition, the color of the RGB tip implies that M110 is at a much higher metallicity than Sculptor.  The isochrones are for $\logz = -0.4$; the 10 Gyr isochrone is only slightly redward of the red giant branch implying a metallicity slightly less than this value.

The measured star formation history indicates that approximately half of the stars in M110 were formed in the past 10 Gyr (compared with under 10\% in the case of Sculptor).  A strong episode of star formation several Gyr ago had approximately an equal star formation rate as the oldest star formation, and is responsible for the red clump.  Because the data do not reach the ancient main sequence turnoff, it is probably unwise to attempt a more detailed analysis of the stars older than a few Gyr.

\subsection{Leo II (DDO 93)}

\begin{figure}[h]
\caption{CMD and star formation history for Leo II. \label{fig_LeoII}}
\end{figure}

While the majority of early-type systems in the Local Group look something like M110 or Sculptor, it is worthwhile to discuss a few more objects in some detail before skimming through the remainder.

The first case is Leo II, which falls towards the fainter end of these systems but has a broad main sequence turnoff indicative of an extended star formation episode.  For the metallicity of the plotted isochrones ($\logz = -1.3$), we find that the color matches the RGB color well, but the turnoff is near the faint end of the observed turnoff.  This implies a relatively large number of stars younger than 10 Gyr.

The star formation history accurately follows these expectations.  We find no significant star formation in the past Gyr, but that slightly over half of the stars were formed within the past 10 Gyr.

\subsection{Carina}

\begin{figure}
\caption{CMD and star formation history for the Carina dSph. From {hk98}, using CTIO 4m observations. \label{fig_Carina}}
\end{figure}

One of the earlier dwarf spheroidals known to have an extended star formation history is the Carina dwarf; its CMD and star formation history are shown in Figure \ref{fig_Carina}.  Rather than a single, broad turnoff as was seen for Leo II, Carina's CMD was fit by \citet{hk98} as the combination of three single-age populations -- approximately 20\% ancient stars, 50\% of the stars approximately 7 Gyr old, and the remaining 30\% of stars formed a few Gyr ago.  Whether the star formation is exactly zero between these populations, or whether the CMD would be equally well fit with a smoother function, is unclear.

\subsection{Leo I (DDO 74)}

\begin{figure}[h]
\caption{CMD and star formation history for Leo I. \label{fig_LeoI}}
\end{figure}

The fifth and final representative early-type system is Leo I, whose CMD and star formation history are shown in Figure \ref{fig_LeoI}.  While Leo I shows the typical ancient population (a turnoff near the oldest isochrone and a horizontal branch), it also shows a main sequence extending slightly youngward of the 1 Gyr isochrone, as well as a vertically-extended red clump that is typical of younger populations.

Examining the star formation history, we find the youngest system yet, with 75\% of its star formation younger than 10 Gyr.  The plot of the star formation history shows a moderate initial star formation rate, rising to a peak approximately 3 Gyr ago and falling to zero by a few hundred Myr ago.

\subsection{Other Early-type Systems}

\subsubsection{M32 (NGC 221)}

\begin{figure}
\caption{CMD and star formation history for M32. \label{fig_M32}}
\end{figure}

The main sequence extends nearly to the 100 Myr isochrone in this system, indicating the presence of a very young population.  Indeed, the average star formation rate over the past Gyr is $\sim 60\%$ of its lifetime average.  As is typical for the more massive early-type systems, about half of the stars were formed in the past 10 Gyr.

\subsubsection{NGC 147 (DDO 3)}

\begin{figure}
\caption{CMD and star formation history for NGC 147. \label{fig_NGC147}}
\end{figure}

Photometric errors are not sufficient to reproduce the large number of blue stars at the base of the CMD.  Therefore the fit is to a star formation history somewhat like that of Leo I, although the recent burst is about 4 Gyr ago.  The shape of the red clump is consistent with this history.

\subsubsection{NGC 185}

\begin{figure}
\caption{CMD and star formation history for NGC 185. \label{fig_NGC185}}
\end{figure}

Similar to NGC 147, except that the ratio of blue HB to red clump stars is larger, and thus the star formation history between the ancient epoch and the most recent large event is flatter.  The detected mini-burst of star formation ~700 Myr ago could be due to older blue stragglers.

\subsubsection{Sagittarius}

\begin{figure}
\caption{CMD and star formation history for the Sagittarius dSph, from \citet{bell99}, using ESO 3.5m NTT observations. \label{fig_Sag}}
\end{figure}

The Sagittarius dwarf is most difficult Galactic companion to study, due to the high amount of foreground contamination.  \citet{bell99} used a statistically cleaned ground-based CMD to determine the history shown here.

\subsubsection{Fornax}

\begin{figure}
\caption{CMD and star formation history for the Fornax dSph, from \citet{sav00}, using observations from the ESO/Danish 1.54m telescope. \label{fig_Fornax}}
\end{figure}

The CMD of the Fornax dwarf shows a main sequence that extends above the horizontal branch, implying the presence of younger stars than are present in most spheroidals.

\subsubsection{Cassiopeia (Andromeda VII)}

\begin{figure}
\caption{CMD and star formation history for the Cassiopeia dSph. \label{fig_NGCCas}}
\end{figure}

The best photometry available for the Cassiopeia dwarf is a pair of 600 second exposures taken as part of a WFPC2 snapshot program.  The handful of bright blue stars may indicate the presence of a young population, or may just be foreground contamination.  Aside from the presence of ancient stars, as evidenced by the blue extension of the HB, little can be said about this system.

\subsubsection{Andromeda I}

\begin{figure}
\caption{CMD and star formation history for Andromeda I. \label{fig_AndI}}
\end{figure}

As with other M31-distance systems, it is unclear whether the brightest blue stars are at the age implied by their main sequence turnoff, or blue stragglers.  The number of stars above the red HB favors the interpretation of a younger population.

\subsubsection{Andromeda II}

\begin{figure}
\caption{CMD and star formation history for Andromeda II. \label{fig_AndII}}
\end{figure}

Similar to Andromeda I, except that the intermediate-age population appears to be slightly stronger.  Approximately 60\% of the star formation occurred less than 10 Gyr ago.

\subsubsection{Andromeda VI}

\begin{figure}
\caption{CMD and star formation history for Andromeda VI. \label{fig_AndVI}}
\end{figure}

Likely an older system ($\sim 25\%$ of the star formation was less than 10 Gyr ago), but the shallower photometry prevents a detailed study of the ancient populations.

\subsubsection{Andromeda III}

\begin{figure}
\caption{CMD and star formation history for Andromeda III. \label{fig_AndIII}}
\end{figure}

The case for a few Gyr old population is much clearer for Andromeda III; there are too many blue stars for all to be blue stragglers.

\subsubsection{Cetus}

\begin{figure}
\caption{CMD and star formation history for the Cetus dSph. \label{fig_Cetus}}
\end{figure}

Another WFPC2 snapshot target, and consequently another highly uncertain history.

\subsubsection{Tucana}

\begin{figure}
\caption{CMD and star formation history for the Tucana dSph. \label{fig_Tucana}}
\end{figure}

Most ($\sim 80\%$) of the stars formed over 10 Gyr ago.  The episode of star formation a few Gyr ago is likely real; the brighter red clump stars and width of the lower RGB indicate this.  However, the 700 Myr burst could be due to residuals of fitting the horizontal branch.

\subsubsection{Sextans}

\begin{figure}
\caption{CMD and star formation history for the Sextans dSph, from \citet{lee03}, using 3.6m CFHT observations. \label{fig_Sextans}}
\end{figure}

A simple, primarily ancient spheroidal.

\subsubsection{Ursa Minor (DDO 199)}

\begin{figure}
\caption{CMD and star formation history for the Ursa Minor dSph, from \citet{car02}, using 2.5m INT observations. \label{fig_UMi}}
\end{figure}

Another simple, primarily ancient spheroidal.

\subsubsection{Andromeda V}

\begin{figure}
\caption{CMD and star formation history for Andromeda V. \label{fig_AndV}}
\end{figure}

The width of the RGB indicates that Andromeda V isn't a simple population like comparably low-luminosity galactic companions (Ursa Minor and Draco).  The nature of any extended star forming event is unknown due to the relatively shallow photometry.

\subsubsection{Draco (DDO 208)}

\begin{figure}
\caption{CMD and star formation history for the Draco dSph, from \citet{apar01}, using 2.5m INT observations. \label{fig_Draco}}
\end{figure}

A simple, primarily ancient spheroidal.  The apparent young burst is almost certainly attributable to blue stragglers.

\subsection{Histories of Spheroidals}

The most noticeable trend among the star formation histories of these systems is that most luminous (and presumably most massive) systems generally had extended episodes of star formation, and formed at least half of their stars during the past 10 Gyr (well more than half for the cases of NGC 147, Sagittarius, and Fornax).  Of the less luminous systems, only Leo I, Andromeda II, Leo II, and Carina show clear signs of a majority of the star formation having come within the past 10 Gyr.

\section{Irregulars}

\subsection{IC 1613 (DDO 8)}

\begin{figure}
\caption{CMD and star formation history for IC 1613. \label{fig_IC1613}}
\end{figure}

A glance at the CMD of IC 1613 (Figure \ref{fig_IC1613}) shows the full set of features that can be present in a color-magnitude diagram: a main sequence extending to the bright limit of the photometry, blue and red helium burning stars, a red giant branch, a red clump, and a horizontal branch.

The recovered star formation history matches these expectations well.  The maximum star formation occurred $\sim 3$ Gyr ago, explaining the overabundance of red clump stars.  At the young end, the ages become degenerate once the main sequence turnoff is brighter than the saturation limit of the photometry (something that will be seen throughout the irregular galaxies), and thus we compute only an average rate over that time.  Otherwise the star formation rate has been constant to within a factor of two; the rate over the past Gyr equals the lifetime average.

An important feature of this system is that the large number of stars between the main sequence and RGB at $I ~ 26$ are not spurious detections near the limits of the photometry.  Instead, those stars represent the main sequence turnoff and subgiant branch of the intermediate-aged stars that constitute the 3 Gyr old star formation event.  The accuracy with which these ages can be probed with deep photometry such as this demonstrates the importance of not settling for less-precise age indicators, such as the red clump or HB morphology, that can be seen in the brighter parts of the CMD.

\subsection{NGC 6822 (DDO 209)}

\begin{figure}[h]
\caption{CMD and star formation history for NGC 6822. \label{fig_NGC6822}}
\end{figure}

The color-magnitude diagram of NGC 6822 (Figure \ref{fig_NGC6822}) looks much like that of IC 1613, except that the youngest populations (most notably, the main sequence) are much stronger, to the extent that the horizontal branch is lost in the noise of the ~1 Gyr turnoff.  (That a horizontal branch exists is known by the presence of RR Lyraes, as reported by \citet{clem03a}.)

Not surprisingly, the measured star formation history is rather similar to that for IC 1613, with the primary difference being the star formation rate over the past 100 Myr.  Normalizing to the lifetime average star formation rates, NGC 6822 has twice as many stars younger than 100 Myr than does IC 1613.  Whether or not this is meaningful is unclear, as both systems show fluctuations in their star formation rates of a factor of two.

\subsection{Leo A}

\begin{figure}[h]
\caption{CMD and star formation history for Leo A. \label{fig_LeoA}}
\end{figure}

Moving from NGC 6822 to Leo A involves a significant reduction in luminosity, and consequently a much less populated CMD (Figure \ref{fig_LeoA}).  The biggest clue that the systems have much different star formation histories comes from the number of blue helium-burning stars (the sequence extending bright and blue from the red clump), which have ages of under 1 Gyr.  The star formation rate during this time has been a factor of two greater than the lifetime average, and a large fraction of Leo A's stars have formed in the past 2 Gyr.

The presence of ancient stars is not immediately apparent from the CMD.  The star formation history measurement program finds a significant ancient population (40\% of the stars are older than 10 Gyr), but difficulties in modeling the oldest populations renders this uncertain.  However, RR Lyrae stars have been detected in Leo A \citep{dolet02}, indicating that a significant number of ancient stars are indeed present and providing some additional confidence in our star formation histories.

\subsection{Sextans A (DDO 75)}

\begin{figure}
\caption{CMD and star formation history for Sextans A. \label{fig_SexA}}
\end{figure}

The final representative irregular galaxy is Sextans A, whose CMD (Figure \ref{fig_SexA}) is dominated even more by young stars than was Leo A.  Sextans A has the highest recent star formation rate (again, normalized to its lifetime average) of any Local Group galaxy -- its star formation rate over the past Gyr is a factor of three times its average, while over the past 100 Myr the ratio is more than five.

As with Leo A, a small number of ancient stars are inferred from the CMD.  However, Sextans A's greater distance has thus far prevented a direct confirmation of this with RR Lyraes.

What caused Sextans A to resume forming stars $2-3$ Gyr ago is unknown, as it is relatively far from any other system in the sky and thus lacks an obvious candidate for interaction.

\subsection{Other Late-type Systems}

\subsubsection{M33 (NGC 598)}

\begin{figure}[h]
\caption{CMD and star formation history for M33. \label{fig_M33}}
\end{figure}

Although not technically a dwarf galaxy, M33 has been included for the sake of completeness as it can be considered a member of the M31 group.  The CMD (Figure \ref{fig_M33}) is a combination of two HST pointings, which gives the impression of two photometric cutoffs ($I \approx 25.5$ and $I = 26.5$).

\subsubsection{LMC}

\begin{figure}
\caption{CMD and star formation history for the LMC.  The history is calculated from a single field, so may not be representative of the LMC's global history.}
\end{figure}

The LMC is covered by the contribution of Harris in this volume.

\subsubsection{IC 10}

\begin{figure}
\caption{CMD and star formation history for IC 10.  The history is calculated from a field at the center of the galaxy, so may not be representative of IC 10's global history.}
\end{figure}

Due to heavy foreground extinction, the IC 10 CMD is extremely shallow.  Thus the caveats from the WFPC2 snapshot targets are applicable here.  Based on the ratio of red to blue stars, the recent star formation rate is approximately a factor of two less than the lifetime average.

\subsubsection{SMC (NGC 292)}

\begin{figure}
\caption{CMD and star formation history for the SMC.  The SFH is a composite of the younger history of \citet{har04} and the older history of \citet{dol01}.}
\end{figure}

The SMC is covered by the contribution of Harris in this volume.

\subsubsection{NGC 3109 (DDO 236)}

\begin{figure}
\caption{CMD and star formation history for NGC 3109.  The history is calculated from a field at the center of the galaxy, so may not be representative of NGC 3109's global history.}
\end{figure}

Another snapshot target with shallow photometry.  As best as can be measured, NGC 3109 has a roughly constant star formation history, except for an increase by a factor of $\sim 2$ in the past 100 Myr.

\subsubsection{IC 5152}

\begin{figure}
\caption{CMD and star formation history for IC 5152.}
\end{figure}

Another snapshot target with shallow photometry.

\subsubsection{WLM (DDO 210)}

\begin{figure}
\caption{CMD and star formation history for WLM. \label{fig_WLM}}
\end{figure}

The CMD and star formation history (Figure \ref{fig_WLM}) are derived from two WFPC2 pointings.  What is significant is that the episodic star formation (with quiescent phases $\sim 4$ and $\sim 1$ Gyr ago were recovered in both solutions, giving reasonable confidence in the recovered history.

\subsubsection{Sextans B (DDO 70)}

\begin{figure}
\caption{CMD and star formation history for Sextans B.}
\end{figure}

Another snapshot target with shallow photometry.

\subsubsection{SagDIG}

\begin{figure}
\caption{CMD and star formation history for SagDIG.}
\end{figure}

The Sagittarius dwarf irregular is an order of magnitude less luminous than any Local Group irregular other than DDO 210.  The small number of stars, moderately high foreground extinction, and shallow snapshot photometry pose quite a challenge for measuring the star formation history.  Nevertheless, the ratio of blue stars to red giants is quite high, so the general shape of our history is fairly robust.

\subsubsection{DDO 210 (Aquarius)}

\begin{figure}
\caption{CMD and star formation history for DDO 210. \label{fig_DDO210}}
\end{figure}

A factor of five less luminous than SagDIG, DDO 210 is by far the least luminous of the known dwarf irregulars.  The low mass has made classification of this system somewhat difficult.  On the basis of its lack of HII regions, this system has generally been classified as a dIrr/dSph galaxy.  However, the CMD (Figure \ref{fig_DDO210}) shows significant numbers of blue stars, resulting in a star formation history that is relatively flat.  The most likely explanation for the lack of HII is that the galaxy's mass and consequently its star formation rate is so low that there don't happen to be any ionizing O stars at present.

\subsection{Histories of Irregulars}

Irregular galaxies seem to have a wide variety of star formation histories.  Although our amount of information is hampered by the shallow photometry available for a large number of these systems, it does seem that the majority of these systems are characterized by episodes of enhanced star formation history separated by lulls.  The exceptions (such as the LMC and IC 1613) are found among the more luminous systems.

Three special systems are Sextans A, the Sagittarius dwarf irregular, and Leo A.  All three have recent star formation rates a factor of two or more higher than their lifetime averages.  It is unclear if these three systems comprise a class of galaxies where star formation was triggered recently, or if they are merely extreme examples of stochastic star formation.

\section{Transition-type Galaxies}

\subsection{Phoenix}

\begin{figure}
\caption{CMD and star formation history for the Phoenix dIrr/dSph. \label{fig_Phoenix}}
\end{figure}

The available photometry for the Phoenix dwarf is the deepest (in absolute magnitude) of any system that is not a Galactic companion, extending to the oldest subgiant stars, and thus giving a robust history for even the ancient populations.  The CMD (Figure \ref{fig_Phoenix}) shows a main sequence extending up to the 100 Myr isochrone, and the full set of older populations.  Particularly prominent is the horizontal branch, something that was often washed out by young stars in the irregular systems.  Indeed, while our measurement finds stars as young as 100 Myr, the majority of star formation in Phoenix occurred more than 10 Gyr ago.

\subsection{Other Transition-type Systems}

\subsubsection{Pegasus (DDO 216)}

\begin{figure}[h]
\caption{CMD and star formation history for the Pegasus dIrr/dSph.}
\end{figure}

The photometry of the Pegasus dwarf isn't nearly as deep, and thus the detailed history at older ages is uncertain.  However, the young stars indicate a rapidly declining star formation rate over the past Gyr, and a recent cessation of star formation.

\subsubsection{Antlia}

\begin{figure}
\caption{CMD and star formation history for the Antlia dIrr/dSph, from Dolphin et al. (in preparation).}
\end{figure}

Of the transition-type galaxies, Antlia shows the greatest amount of recent star formation, most notable in the CMD by the blue helium-burning sequence.  However, as with the others, it shows a strong red horizontal branch, indicating a high star formation rate at ancient times.

\subsubsection{LGS 3 (Pisces)}

\begin{figure}
\caption{CMD and star formation history for LGS 3.}
\end{figure}

Virtually a clone of Phoenix.

\begin{table}[!ht]
\caption{Star formation history properties of Local Group galaxies \label{tab_SFH}}
\smallskip
\begin{center}
{\small
\begin{tabular}{lccccccc}
\tableline
\noalign{\smallskip}
Name & Type & $\dist$\tablenotemark{a} & $A_V$\tablenotemark{b} & $b_{100M}$$^c$ & $b_{1G}$$^d$ & $f_{10G}$$^e$ & $\mean{\logz}^f$ \\
\noalign{\smallskip}
\tableline
\noalign{\smallskip}
WLM         & Irr       & $24.85 \pm 0.12$   & 0.123 & 0.5 & 1.2 & 0.68 & -1.0 \\
IC 10       & Irr       & $24.53 \pm 0.13$ & 2.38$^g$ & 0.6 & 0.3 & 0.56 & -0.8 \\
Cetus       & dSph      & $24.44 \pm 0.11$   & 0.095 & 0.0 & 0.2 & 0.29 & -1.4 \\
NGC 147     & dE5       & $24.28 \pm 0.13$   & 0.574 & 0.0 & 0.1 & 0.64 & -0.9 \\
And III     & dSph      & $24.29 \pm 0.15^h$ & 0.187 & 0.0 & 0.0 & 0.40 & -1.3 \\
NGC 185     & dE3p      & $23.94 \pm 0.12$   & 0.605 & 0.0 & 0.1 & 0.49 & -0.9 \\
M110        & E5p       & $24.45 \pm 0.14$   & 0.206 & 0.0 & 0.1 & 0.54 & -0.6 \\
M32         & E2        & $24.47 \pm 0.15$   & 0.206 & 0.0 & 0.6 & 0.50 & -0.3 \\
And I       & dSph      & $24.42 \pm 0.16^h$ & 0.180 & 0.0 & 0.0 & 0.42 & -1.1 \\
SMC         & Irr       & $18.91 \pm 0.11^i$ & 0.123 & 1.7 & 1.3 & 0.69 & -1.1 \\
Sculptor    & dSph      & $19.47 \pm 0.17^i$ & 0.059 & 0.0 & 0.0 & 0.05 & -1.8 \\
LGS 3       & dIrr/dSph & $23.92 \pm 0.15^h$ & 0.136 & 0.0 & 0.2 & 0.36 & -1.3 \\
IC 1613     & Irr       & $24.36 \pm 0.12$   & 0.083 & 0.6 & 1.0 & 0.69 & -1.1 \\
And V       & dSph      & $24.46 \pm 0.15^h$ & 0.413 & 0.0 & 0.0 & 0.35 & -1.4 \\
And II      & dSph      & $24.06 \pm 0.16^h$ & 0.207 & 0.0 & 0.0 & 0.60 & -1.2 \\
M33         & Sc        & $24.73 \pm 0.12$   & 0.139 & 0.7 & 1.3 & 0.52 & -0.7 \\
Phoenix     & dIrr/dSph & $23.07 \pm 0.15^h$ & 0.052 & 0.0 & 0.4 & 0.42 & -1.3 \\
Fornax      & dSph      & $20.66 \pm 0.15^i$ & 0.067 & 0.0 & 0.0 & 0.73 & -1.0 \\
LMC         & Irr       & $18.52 \pm 0.09^i$ & 0.249 & 1.1 & 1.5 & 0.82 & -0.7 \\
Carina      & dSph      & $20.09 \pm 0.12^i$ & 0.208 & 0.0 & 0.0 & 0.80 & -2.0 \\
Leo A       & dIrr      & $24.49 \pm 0.14$   & 0.068 & 1.7 & 2.1 & 0.61 & -1.4 \\
Sextans B   & dIrr      & $25.70 \pm 0.12$   & 0.105 & 0.9 & 0.8 & 0.62 & -1.2 \\
NGC 3109    & Irr       & $25.62 \pm 0.14$   & 0.221 & 1.4 & 0.8 & 0.61 & -1.1 \\
Antlia      & dIrr/dSph & $25.47 \pm 0.14$   & 0.263 & 0.4 & 0.6 & 0.43 & -1.0 \\
Leo I       & dSph      & $22.03 \pm 0.14$   & 0.120 & 0.0 & 0.3 & 0.75 & -1.2 \\
Sextans A   & dIrr      & $25.63 \pm 0.12$   & 0.145 & 5.3 & 3.0 & 0.50 & -1.4 \\
Sextans     & dSph      & $19.67 \pm 0.15^i$ & 0.165 & 0.0 & 0.0 & 0.00 & -2.1 \\
Leo II      & dSph      & $21.48 \pm 0.16^h$ & 0.055 & 0.0 & 0.0 & 0.56 & -1.3 \\
Ursa Minor  & dSph      & $19.41 \pm 0.12^i$ & 0.105 & 0.0 & 0.0 & 0.00 & -1.9 \\
Draco       & dSph      & $19.84 \pm 0.14^i$ & 0.091 & 0.0 & 0.0 & 0.06 & -1.8 \\
Sagittarius & dSph      & $17.10 \pm 0.15^i$ & 0.508 & 0.0 & 0.0 & 0.84 & -1.2 \\
SagDIG      & dIrr      & $25.24 \pm 0.17$   & 0.401 & 2.9 & 1.4 & 0.64 & -1.3 \\
NGC 6822    & Irr       & $23.60 \pm 0.12$   & 0.784 & 1.3 & 1.1 & 0.69 & -1.0 \\
DDO 210     & dIrr      & $24.86 \pm 0.14$   & 0.170 & 0.8 & 0.9 & 0.62 & -1.7 \\
IC 5152     & Irr       & $26.39 \pm 0.19$   & 0.082 & 1.3 & 0.7 & 0.61 & -0.8 \\
Tucana      & dSph      & $24.66 \pm 0.13^h$ & 0.105 & 0.0 & 0.1 & 0.22 & -1.6 \\
Cassiopeia  & dSph      & $24.54 \pm 0.13$   & 0.644 & 0.0 & 0.1 & 0.28 & -1.1 \\
Pegasus     & dIrr/dSph & $24.88 \pm 0.17$   & 0.218 & 0.2 & 0.8 & 0.46 & -0.9 \\
And VI      & dSph      & $24.60 \pm 0.15^h$ & 0.212 & 0.0 & 0.0 & 0.22 & -0.9 \\
\noalign{\smallskip}
\tableline
\noalign{\smallskip}
\tablenotetext{a}{Measured from the RGB tip (adopting $M_I = -4.01 \pm 0.10$ for metal-poor stars) unless otherwise noted.}
\tablenotetext{b}{From \citet{sch98} unless otherwise noted.}
\end{tabular}
}
\end{center}
\end{table}

\begin{table}
Table 1. (continued)
\smallskip
\begin{center}
{\small
\begin{tabular}{l}
\tablenotetext{c}{The average star formation rate over the past 100 Myr, normalized to the galaxy's lifetime average.}
\tablenotetext{d}{The average star formation rate over the past Gyr, normalized to the galaxy's lifetime average.}
\tablenotetext{e}{The fraction of the galaxy's star formation that has occured in the past 10 Gyr.}
\tablenotetext{f}{From the star formation histories.  We note that these values tend to exceed spectroscopically-measured metallicities.}
\tablenotetext{g}{Extinction obtained from CMD fitting.}
\tablenotetext{h}{Distance computed from the horizontal branch, adopting $M_V = -0.61 \pm 0.14$.}
\tablenotetext{i}{
Carina: \citet{dall03};
Draco and Ursa Minor: \citet{bell02};
Fornax: \citet{mack03};
LMC: \citet{clem03b};
Sagittarius: \citet{mon04};
Sculptor: based on RR Lyraes of \citet{kal95};
Sextans: \citet{mat95};
SMC: \citet{hild05}.}
\end{tabular}
}
\end{center}
\end{table}

\subsection{Histories of Transition-type Systems}

The four transition-type galaxies show similar star formation histories.  All formed half or more of their stars more than 10 Gyr ago, and have recent (past 100 Myr) star formation rates that are less than half of their lifetime averages.  It appears that these systems are former irregulars that have exhausted most of their gas and will evolve into spheroidals in the near future.

\section{Summary}

Recent improvements in stellar evolution models and computing power have permitted star formation histories to be measured with unprecedented detail.  Rather than measuring histories very roughly based on counts of stars in various evolutionary phases, modern techniques fit all features of the CMD using statistical comparisons between observed and synthetic CMDs.  This permits not only a more detailed measurement of the star formation history, but also rough metallicity information.

We have applied these techniques to a database containing deep photometry of most Local Group galaxies, producing the first homogeneous analysis of the Local Group using current star formation history techniques.  The resulting star formation histories are summarized in Figures \ref{fig_SFH_sph} and \ref{fig_SFH_irr} and Table \ref{tab_SFH}.

\begin{figure}
\caption{Star formation histories of elliptical and spheroidal galaxies.  Colors correspond to the CMD features generated by each age.  Red: RGB plus full HB. Orange: RGB plus red HB.  Yellow: RGB plus red clump. Green: bright red clump. Blue: young MS and blue helium-burning stars. \label{fig_SFH_sph}}
\end{figure}

\begin{figure}
\caption{Star formation histories of irregular and transition-type galaxies.  Colors correspond to the CMD features generated by each age.  Red: RGB plus full HB. Orange: RGB plus red HB.  Yellow: RGB plus red clump. Green: bright red clump. Blue: young MS and blue helium-burning stars. \label{fig_SFH_irr}}
\end{figure}

Among the spheroidals and ellipticals, we find a correlation between absolute magnitude and the duration of the epoch of star formation, in the sense that more luminous galaxies generally formed their stars over a larger period of time.  Specifically, all but four (Leo I, Andromeda II, Leo II, and Carina) of the fifteen least-luminous spheroidals formed the majority of their stars more than 10 Gyr ago, while all of the six most-luminous ellipticals and spheroidals formed at least half of their stars within the past 10 Gyr.

The irregulars show another trend with absolute magnitude, as the most luminous systems have the most steady star formation rates.  This is not surprising, as one would expect the fluctuations due to stochastic star formation to have a greater effect in smaller systems.

Comparing the histories of the various morphological types, we find that a galaxy's morphological type is entirely determined by its star formation history during the past Gyr or less.  No spheroidal or elliptical galaxy has a significant amount of star formation in the past 100 Myr, and only M32 and Leo I show a significant amount of star formation in the past Gyr.  In contrast, every irregular galaxy has a recent star formation rate (defined as the mean rate over the past 100 Myr) of at least half its lifetime average, and all have ongoing star formation.  As expected, transition-type galaxies fall in between, with minimal ongoing star formation and a recent star formation rate of between zero and half the lifetime average.

The morphological classifications are not always unambiguous, with Antlia's properties placing it near the irregular/transition-type boundary while Leo I and M32 fall close to the speroidal/transition-type boundary.  In addition, all systems observed in sufficient detail have shown evidence of ancient populations -- even Leo A and Sextans A, which are the most strongly dominated by young populations.  The implication is that there is no firm division between the dwarf galaxy morphologies, but rather an evolutionary sequence: once a galaxy exhausts (or otherwise loses) its gas, it will change from an irregular through the transition-type phase, and eventually become a spheroidal.

\acknowledgements             

Support for Proposal number 9521 was provided by NASA through a grant from the Space Telescope Science Institute, which is operated by the Association of Universityies for Research in Astronomy, Incorporated, under NASA contract NAS5-26555.


\end{document}